\documentclass{aa}

\usepackage[]{graphicx}
\usepackage[]{graphics}
\usepackage[]{psfig}
\usepackage[]{epsfig}
\usepackage{txfonts}
\usepackage[]{longtable}
\usepackage[]{times}
\include{epsf}

\begin{document}
\input{psfig.sty}


\title{Dynamical masses of ultra-compact dwarf galaxies in Fornax\thanks{Based 
on observations obtained at the European Southern Observatory, Chile (Observing
Programme 66.B--0378).}}

\author {Michael Hilker \inst{1,2} \and Holger Baumgardt \inst{1} \and Leopoldo 
Infante \inst{3} \and Michael Drinkwater \inst{4} \and Ekaterina Evstigneeva 
\inst{4} \and Michael Gregg \inst{5,6}}

\offprints {M.~Hilker}
\mail{mhilker@eso.org}

\institute{
Argelander-Institut f\"ur Astronomie, Universit\"at Bonn, Auf dem H\"ugel 71, 
53121 Bonn, Germany\thanks{Founded by merging of the Sternwarte, 
Radioastronomisches Institut and Institut f\"ur Astrophysik und 
Extraterrestrische Forschung der Universit\"at Bonn}
\and
European Southern Observatory, Karl-Schwarzschild-Str.~2, 85748 Garching bei
M\"unchen, Germany
\and
Departamento de Astronom\'\i a y Astrof\'\i sica, P.~Universidad Cat\'olica,
Casilla 306, Santiago 22, Chile
\and
Department of Physics, University of Queensland, Brisbane, QLD 4072, Australia
\and
Department of Physics, University of California, Davis, 1 Shields Avenue, 
Davis, CA 95616, USA
\and
Institute of Geophysics and Planetary Physics, L-413, Lawrence Livermore 
National Laboratory, Livermore, CA 94550, USA
}

\date{Received / Accepted }

\titlerunning{Dynamical masses of of ultra-compact dwarf galaxies in Fornax}

\authorrunning{M.~Hilker et al.}

\abstract
{}
{We determine masses and mass-to-light ratios of five ultra-compact dwarf 
galaxies and one dwarf elliptical nucleus in the Fornax cluster from 
high resolution spectroscopy. We examine whether they are consistent with
pure stellar populations or whether dark matter is needed to explain their
masses.} 
{Velocity dispersions were derived from selected wavelength regions using a 
direct-fitting method. To estimate the masses of the UCDs a new 
modelling program has been developed that allows a choice of different 
representations of the surface brightness profile (i.e. Nuker, Sersic or King 
laws) and corrects the observed velocity dispersions for observational 
parameters (i.e. seeing, slit size). The derived dynamical masses are
compared to those expected from stellar population models.}
{The observed velocity dispersions range between 22 and 30 km\,s$^{-1}$.
The resulting masses are between 1.8 and $9.5\times10^7M_{\odot}$. These, as
well as the central and global projected velocity dispersions, were derived
from the generalized King model which turned out to give the most stable 
results. The masses of two UCDs, that are best fitted by a two-component 
profile, were derived from a combined King+Sersic model. The mass-to-light 
ratios of the Fornax UCDs range between 3 and 5 $(M/L_V)_{\odot}$. The $M/L_V$ 
ratio of the dwarf elliptical nucleus is 2.5.
These values are compatible with predictions from stellar population 
models. Within 1-2 half-mass radii dark matter is not dominating UCDs
and the nucleus. An increasing dark matter contribution towards larger
radii can not be ruled out with the present data. The $M/L_V$ ratios of 
some UCDs suggest they have intermediate age stellar populations.}
{We show that the mass-to-light ratios of UCDs in Fornax are
consistent with those expected for pure stellar populations. Thus UCDs
seem to be the result of cluster formation processes within galaxies rather 
than being compact dark matter dominated substructures themselves. Whether 
UCDs gained their mass in super-star
cluster complexes of mergers or in nuclear star cluster formation
processes remains an open question. It appears, however, clear that star
clusters more massive than about $5\times10^6M_{\odot}$ exhibit a
more complex formation history than the less massive `ordinary' globular
clusters.}


\maketitle

\section{Introduction}
\label{intro}

In the late 1990s a new type of very bright compact stellar system was
discovered in the core of the nearby Fornax galaxy cluster (Hilker et al.
\cite{hilk99c}, Drinkwater et al. \cite{drin00a}). On ground-based images
with typical resolution ($\sim0.8$ arcsec seeing) these objects look like
stars since at Fornax distance globular clusters (GCs) are unresolved.
However, with absolute 
magnitudes in the range $M_V=-13.5$ to $-12.3$ they are about 2 mag brighter 
than $\omega$ Centauri, the most massive globular cluster in the Milky Way, 
but about 3 mag fainter than the compact dwarf elliptical M32. Their 
luminosities are comparable to those of nuclei of dwarf ellipticals or 
late-type spirals (i.e. Lotz et al. \cite{lotz04}, Walcher et al. 
\cite{walc05}, C\^ot\'e et al. \cite{cote06}). Their sizes range between 10
and 30 pc in half-light radii (Drinkwater et al. \cite{drin03a}) which is 
about 10 times larger than ordinary GCs. 

\begin{table*}
\caption{\label{tab1} Observation log and properties of observed objects.
$V_0$ is from Evstigneeva et al. (\cite{evst06}) and $(V-I)_0$ from Mieske et
al. (\cite{mies06}) (except $(V-I)_0$ for FCC303 which is from Karick et al.
\cite{kari03}). Both, $V_0$ and $(V-I)_0$ are corrected for foreground dust
extinction (Schlegel et al.  \cite{schl98}). The [Fe/H] values are from Mieske
et al. (\cite{mies06}).}
\begin{tabular}{p{0.9cm}p{2.5cm}ccp{0.6cm}p{0.9cm}@{\hspace{2mm}}p{0.6cm}lccp{0.6cm}p{0.6cm}}
\hline
 & & & & & & & & & & \\[-3mm]
Name & FCSS$^a$ name & $\alpha$(2000) & $\delta$(2000) & $V_0$ & $(V-I)_0$ &
 [Fe/H] & Date & Exp.~time & Seeing & S/N$^b$ & S/N$^b$ \\
 & & [h:m:s] & [$^\circ$:$\arcmin$:$\arcsec$] & [mag] & [mag] & [dex] & & [s] &
[$\arcsec$] & 4900$\AA$ & 5900$\AA$ \\
\hline
 & & & & & & & & & & \\[-3mm]
FCC303$^c$ &              & 3:45:14.10 & $-$36:56:12.0 & 15.90 & 1.09 & &
2000~Dec~20 & $3\times1004$ & 0.80 & 3.6 & 4.2 \\
 & & & & & & & & & \\[-3mm]
UCD2 & J033806.3$-$352858 & 3:38:06.33 & $-$35:28:58.8 & 19.12 & 1.14 & $-$0.90
 & 2000~Dec~23 & $5\times1800$ & 0.55 & 4.5 & 5.6 \\
 & & & & & & & & & \\[-3mm]
UCD3 & J033854.1$-$353333 & 3:38:54.10 & $-$35:33:33.6 & 17.82 & 1.22 & $-$0.52
 & 2000~Dec~20 & $3\times1000$ & 0.80 & 4.7 & 5.9 \\
 & & & & & & & & & \\[-3mm]
UCD4 & J033935.9$-$352824 & 3:39:35.95 & $-$35:28:24.5 & 18.94 & 1.13 & $-$0.85
 & 2000~Dec~24 & $5\times1800$ & 0.60 & 4.8 & 5.6 \\
 & & & & & & & & & \\[-3mm]
UCD5 & J033952.5$-$350424 & 3:39:52.58 & $-$35:04:24.1 & 19.40 & 1.00 & &
2001~Feb~12 & $2\times3285$ & 0.60 & 4.4 & 5.2 \\
UCD5 & & & & & & & 2001~Feb~13 & $2\times3285$ & 0.55 & 4.8 & 5.8 \\
UCD5 & & & & & & & 2001~Aug~19 & $1\times3285$ & 0.55 & 4.6 & 5.1 \\
UCD5 & & & & & & & 2001~Sep~20 & $1\times3119$ & 0.80 & 5.3 & 6.3 \\
UCD5 & & & & & & & 2001~Nov~19 & $1\times3285$ & 0.85 & 5.2 & 6.4 \\
\hline
\end{tabular}
\vskip0.1cm
$^a$Fornax Cluster Spectroscopic Survey (Drinkwater et al. \cite{drin99});
$^b$measured for a single spectrum per resolution element ($= 0.03$\AA);
$^c$Fornax Cluster Catalog (Ferguson \cite{ferg89a})
\end{table*}

In the last years, extensive surveys in the Fornax and Virgo cluster have
revealed dozens of these compact objects (UCDs and bright GCs) in the 
magnitude range
$-13.5<M_V<-11.0$ (i.e. Mieske et al. \cite{mies02} and \cite{mies04a},
Ha\c{s}egan et~al. \cite{hase05}, Jones et al. \cite{jonej06}). They were
dubbed ''ultra-compact dwarf galaxies'' (UCDs) by Phillipps et al.
(\cite{phil01}), and ''dwarf-globular transition objects'' (DGTOs) by
Ha\c{s}egan et~al. (\cite{hase05}), expressing their uncertain origin.
Various formation scenarios have been suggested to explain the origin
of UCDs. The most promising are:\\

1) UCDs are the remnant nuclei of galaxies that have been significantly
stripped in the cluster environment (i.e. Bassino et al. \cite{bass94},
Bekki et al. \cite{bekk03b}). This would mean that UCDs once resided in the
centres of dark matter (DM) dominated cosmological substructures. The isolated
nuclei themselves, however, are probably not DM dominated since DM halos of
galaxies can be stripped (Bekki et al. \cite{bekk03b}). Moreover, present-day
nuclei of dwarf ellipticals are dominated by their stellar populations, not
by dark matter (Geha et al. \cite{geha02}).

\vskip 2mm
2) UCDs formed from the agglomeration of many young, massive star clusters
that were formed during merger events (i.e. Kroupa \cite{krou98},
Fellhauer \& Kroupa \cite{fell02a}). Hence, UCDs would be dark matter free
stellar objects that were not created in the centres of galaxies.

\vskip 2mm
3) UCDs are the brightest globular clusters formed in the same GC formation
event as their less massive counterparts (i.e. Mieske et al. \cite{mies02},
Martini \& Ho \cite{martin04}). Also in this case, UCDs would be objects
possessing pure stellar populations.

\vskip 2mm
4) UCDs are the surviving counterparts of genuine compact dwarf galaxies that 
formed in small dark matter halos at the low mass end of cosmological 
sub-structure (i.e. Drinkwater et al. \cite{drin04}). Nowadays UCDs survived
the galaxy cluster formation and 
evolution processes, e.g. they were not disrupted or accreted by larger 
galaxies until the present time. This would mean that they might still be 
dominated by dark matter.\\

To distinguish the different formation scenarios one needs to study
the detailed properties of UCDs, like structural parameters, internal
kinematics, together with ages and chemical compositions of their stellar
population(s). In particular, the internal velocity dispersion $\sigma_0$ of
the stars within the UCDs is of interest, from which masses and mass-to-light
($M/L$) ratios can be estimated. A high virial $M/L$ ratio would imply a 
compact cosmological progenitor rather than a merged stellar super-cluster.

Recent high spatial resolution imaging and mid- to high-resolution
spectroscopy shed some light on the properties of UCDs.
First results on the velocity dispersion of UCDs in Fornax were presented in
Drinkwater et al. (\cite{drin03a}). These were based on a quick analysis of
the same high resolution spectra that are the basis for this paper. The
observed, projected velocity dispersions ranged between 24 and 37 km\,s$^{-1}$.
Using the sizes of the UCDs -- derived from high resolution surface brightness
profiles -- their masses and mass-to-light ratios ($M/L_V$) were roughly
estimated. $M/L_V$ values were of the order 2-4 in solar units. This is
slightly higher than the $M/L$ ratio of globular clusters, but much lower than
that of dwarf spheroidal galaxies of similar mass. In the present paper, we 
revise the formerly obtained masses and $M/L$ ratios of the Fornax UCDs.

In the Virgo cluster, Ha\c{s}egan et~al. (\cite{hase05}) derived masses and
$M/L$ ratios for six DGTOs/UCDs. Three of those have relatively high $M/L_V$
values between 6 and 9 at velocity dispersions between 20 and 30 km\,s$^{-1}$.
This makes them the first UCDs with some evidence for dark matter, given their
rather low metallicities ($-1.5<$[Fe/H]$<-0.7$ dex).
Fellhauer \& Kroupa (\cite{fell06}), however, point out that the tidal
``heating'' due to a close passage of an UCD through the central region of
its host galaxy (pericenter between 100 and 1000pc) may lead to an
overestimation of its virial mass {\it at all orbital radii} due to the
presence of unbound stars in the line of sight. The effect would occur mostly
for UCDs with nearly radial orbits. Such orbits are indeed predicted for
the remnant nuclei of stripped galaxies (Bekki et al. \cite{bekk03b}) because
of the selective way tidal stripping works.

\begin{figure}
\psfig{figure=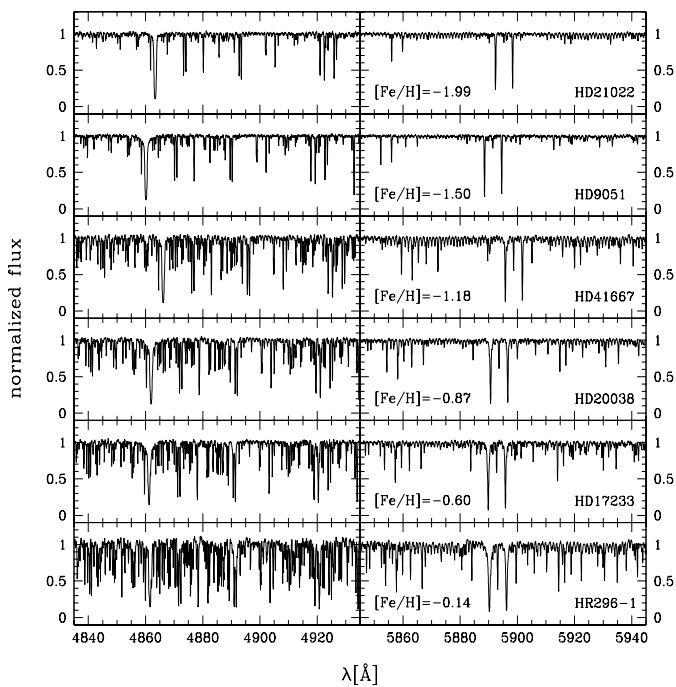,height=8.6cm,width=8.6cm
,bbllx=9mm,bblly=65mm,bburx=201mm,bbury=244mm}
\vspace{0.4cm}
\caption{\label{fig1} The six template spectra (see Table~\ref{tab2}) are shown
in 100$\AA$ wide regions around the H$\beta$ line ($\sim4860\AA$, left panels)
and the Na\,{\sc I} doublet ($\sim5890\AA$, right panels). The spectra are
ordered by increasing metallicity as indicated.}
\end{figure}

\begin{figure}
\psfig{figure=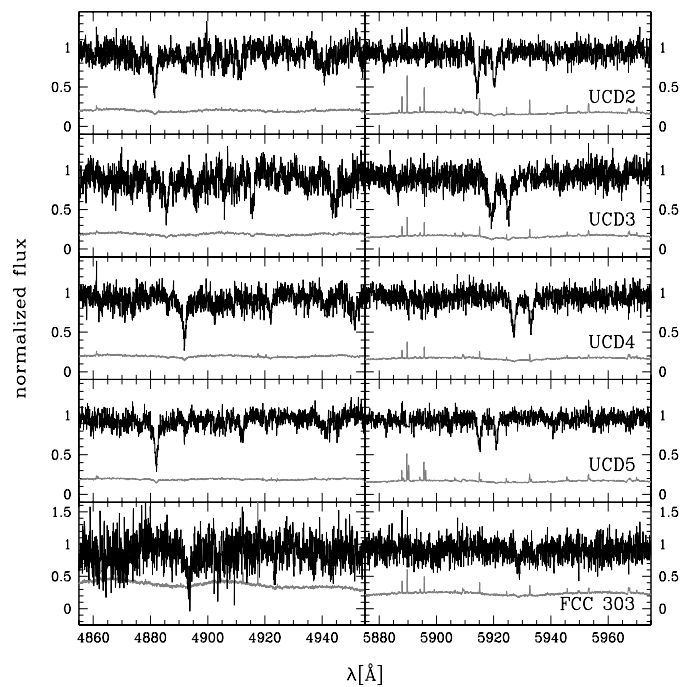,height=8.6cm,width=8.6cm
,bbllx=9mm,bblly=65mm,bburx=201mm,bbury=244mm}
\vspace{0.4cm}
\caption{\label{fig2} The five object spectra (see Table~\ref{tab1}) are shown
in 100$\AA$ wide regions around the H$\beta$ line ($\sim4885\AA$, left panels)
and the Na\,{\sc I} doublet ($\sim5920\AA$, right panels). The 1-sigma error
are plotted as well (grey lines).}
\end{figure}

\begin{table*}
\caption{\label{tab2} Observation log and properties of standard stars that
were used as  velocity dispersion templates. They are ordered by their
increasing metallicity. Most of the properties were taken from the SIMBAD
database, see: {\tt http://simbad.u-strsbg.fr/Simbad}}
\begin{tabular}[l]{l@{\hspace{2mm}}l@{\hspace{2mm}}ccccccccccc}
\hline
 & & & & & & & & & & & & \\[-3mm]
\# & Name & Type$^a$ & $\alpha$(2000) & $\delta$(2000) & $V$ & $(B-V)$ & [Fe/H]
 & $v_{\rm helio}$ & Date & Exp. & S/N$^b$ & S/N \\
 & & & [h:m:s] & [$^\circ$:$\arcmin$:$\arcsec$] & [mag] & [mag] & [dex] &
[km s$^{-1}$] & & [s] & 4900$\AA$ & 5900$\AA$ \\
\hline
 & & & & & & & & & & & & \\[-3mm]
1 & HD~21022 & G & 03:22:21.59 & -32:59:39.7 & 9.20 & 0.95 & $-$1.99 &
 110$\pm$3 & 2000~Dec~20 & 50 & 50.0 & 55.0 \\
2 & HD~9051 & G & 01:28:46.47 & -24:20:25.3 & 8.93 & 0.82 & $-$1.50 &
 $-$73$\pm$1 & 2000~Dec~20 & 40 & 52.0 & 57.0 \\
3 & HD~41667 & G & 06:05:03.64 & -32:59:39.3 & 8.53 & 1.01 & $-$1.18 &
 302$\pm$10 & 2000~Dec~23 & 30 & 49.0 & 53.0 \\
4 & HD~20038 & G & 03:10:26.81 & -58:49:40.5 & 8.90 & 0.84 & $-$0.87 &
 27$\pm$10 & 2000~Dec~24 & 40 & 52.0 & 55.0 \\
5 & HD~17233 & SG & 02:43:52.03 & -54:47:28.3 & 9.03 & 0.79 & $-$0.60 &
 $-$18$\pm$10 & 2001~Feb~13 & 45 & 49.0 & 53.0 \\
6 & HR~296-1 & K0III & 01:03:02.50 & -04:50:12.0 & 5.43 & 1.11 & $-$0.14 &
 15$\pm$? & 2001~Aug~19 & 5 & 58.0 & 61.0 \\
7 & HR~296-2 & & & & & & & & 2001~Sep~20 & 5 & 59.0 & 62.0 \\
8 & HR~296-3 & & & & & & & & 2001~Nov~19 & 2.5 & 53.0 & 57.0 \\
\hline
\end{tabular}
\vskip0.1cm
$^a$G: giant; SG: supergiant; $^b$measured for a single spectrum per resolution
element ($= 0.03\AA$)
\end{table*}

The structural parameters of the brightest Virgo and Fornax UCDs were
presented in several recent papers (i.e. Ha\c{s}egan et~al. \cite{hase05},
De Propris et al. \cite{depr05}, Evstigneeva et al. \cite{evst06}).
A striking result is that UCDs brighter than $M_V=-11.5$ follow a
luminosity-size relation (Ha\c{s}egan et~al. \cite{hase05}, Richtler
\cite{rich06}, Kissler-Patig et al. \cite{kiss06}, Mieske et al.
\cite{mies06}), different from the luminosity independent sizes of globular
clusters (i.e. Jord\'an et al. \cite{jord05}). Moreover, the light profiles
of the brightest
UCDs cannot be described by a simple King profile (Evstigneeva et al.
\cite{evst06}). Most of them show a cuspy profile in their centres and no
tidal truncation within the observed surface brightness limit. Some of them
exhibit small envelopes with effective radii upto 100 pc (e.g. Drinkwater et
al. \cite{drin03a}; Richtler et al. \cite{rich05}).

Concerning the chemical abundances of UCDs, recent work suggests that most
of the bright Virgo UCDs are old and have metallicities ranging from
[Z/H]$=-1.35$ to $+0.35$ dex with an [$\alpha$/Fe] value of $\simeq$0.3
(Evstigneeva et al. \cite{evst06}), whereas the brightest Fornax UCDs are
predominantly metal-rich ([Fe/H]$\simeq-0.5\pm0.5$ dex; Mieske et al.
\cite{mies06}). Their ages and $\alpha$-abundances are not constrained very
well. Future surveys will clarify whether some UCDs may exhibit
intermediate-age stellar populations.

\begin{figure*} 
\psfig{figure=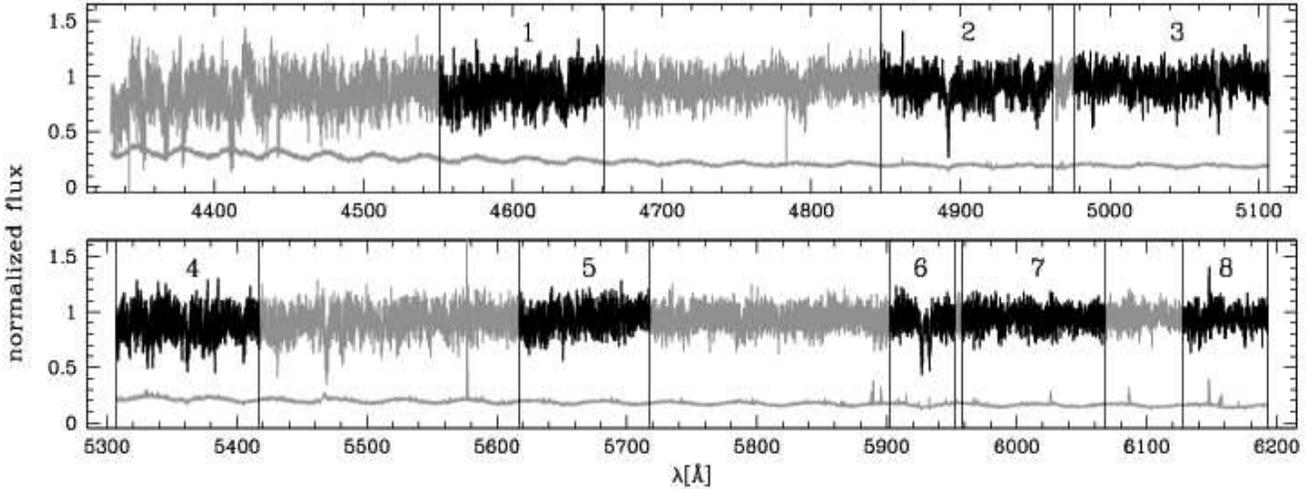,height=6.3cm,width=17.7cm
,bbllx=16mm,bblly=134mm,bburx=197mm,bbury=194mm}
\vspace{0.4cm}
\caption{\label{fig3} The ``blue'' part (upper panel) and the ``red'' part
(lower panel) of the spectrum of UCD4 are shown. Dark regions mark the eight
direct fitting regions for velocity dispersion measurements (see
Table~\ref{tab3}). The grey wavy lines are the 1-sigma error spectra.
}
\end{figure*}

In this paper, we determine dynamical masses of five of the
brightest UCDs and one nucleus of a dwarf elliptical in the Fornax cluster.
The work is based on the detailed analysis of high resolution spectroscopy
that was obtained with the Very Large Telescope Ultraviolet and Visual Echelle
Spectrograph (VLT/UVES)
at Paranal/Chile (and Keck spectroscopy for one of the UCDs). The derivation
of structural parameters for the UCDs --
which are necessary to estimate virial masses -- are taken from
Evstigneeva et al. (\cite{evst06}). In section~2 of the
present paper, the observations and data processing are described.
Section~3 is dedicated to the velocity dispersion measurements, and section~4
to the derivation of masses and mass-to-light ratios by dynamical modelling
and their comparison with expected values from stellar population models.
Finally, in section~5 the main results are summarized.
Throughout the paper we use a distance modulus of 31.39 mag for the Fornax
cluster (Freedman et al. \cite{freed01}).

\section{Observations and data reduction}

\subsection{Observations}

The observations were performed between December 2000 and November 2001 with
the VLT/UT2 at Paranal (ESO, Chile) in service mode. The instrument in use was
the Ultraviolet and Visual Echelle Spectrograph (UVES), and an array of two
attached 2$\times$4\,k CCD chips. The data were read out with a 2$\times$2
binning, resulting in a spatial scale of 0.36\arcsec/pixel. The setting used
(RED arm, $\lambda_{\rm central}=5200\AA$) provided a wavelength coverage of
4200-6200$\AA$, with a gap between 5160 and 5240$\AA$ (blind region beween the
two CCD chips). The slit was chosen to have a width of 1\arcsec which gives
a resolving power of 37000-45000. The final resolution ranged between 2.4 and
3.1 pixels (or 0.12 and 0.16$\AA$). It depends on the seeing which varied
between $0\farcs5$ and $0\farcs9$.

\begin{table}
\caption{\label{tab3} Wavelength regions for direct fitting of velocity
dispersions and prominent lines therein. The values are given in the
restframe.}
\begin{tabular}{llll}
\hline
 & & & \\[-3mm]
Id & $\lambda_{\rm start}$ & $\lambda_{\rm end}$ & Absorption lines \\
 & [$\AA$] & [$\AA$] & \\
\hline
 & & & \\[-3mm]
1 & 4523.2 & 4632.7 & Mg{\sc I}$_{4571}$ \\
2 & 4816.8 & 4931.3 & C{\sc IV}, H$\beta$ \\
3 & 4946.2 & 5075.6 & Fe$_{5041}$ \\
4 & 5274.6 & 5384.1 & Fe$_{5371}$ \\
5 & 5583.1 & 5682.6 & \\
6 & 5866.8 & 5916.5 & Na{\sc I}~(D2), Na{\sc I}~(D1)\\
7 & 5921.5 & 6031.0 & \\
8 & 6090.7 & 6155.4 & Ca{\sc I}1, Ca{\sc I}2 \\
\hline
\end{tabular}
\vskip0.1cm
\end{table}

In total, four ultra-compact dwarfs in the magnitude range $17.8 < V < 19.4$
mag and one nucleus of a dwarf elliptical ($V_{\rm nuc} \simeq 19.5$) were
observed. Most of the observations were taken in three nights in December
2000. Only for one object (UCD5), the single exposures were taken in five
different nights between February and November 2001. The total integration time
for the UCDs varied between 50 minutes and 6 hours. Table~\ref{tab1} gives
a short logbook of the observations, together with the achieved S/N at two
different wavelengths (per 0.05$\AA$ pixel). Also the properties of the
observed objects are summarized in this table.

In the same run, spectra of six standard stars were taken with the same
instrument configuration. They mostly are giants and cover a metallicity range
of $-2.0<$[Fe/H]$<-0.1$ dex. Their properties are summarized in
Table~\ref{tab2}, and excerpts of their spectra around the H$\beta$ line and
the Na doublet are shown in Fig.~\ref{fig1}. The same spectral regions (however
redshifted) for the UCDs and FCC303 are shown in Fig.~\ref{fig2}.

\subsection{Data reduction}

The European Southern Observatory (ESO) provides pipeline reduced data for
UVES echelle spectroscopy. The UVES context in the MIDAS environment is used
for that reduction. Due the low S/N of our data, the pipeline reduced
data with standard settings can not be trusted right away. Therefore, we 
decided to re-reduce these data with the same pipeline, carefully controlling 
all steps and optimizing the setup parameters. Also, we implemented a 
different more reliable cosmic ray removal routine as the one used by the 
pipeline.

The first steps, namely first guess solution, defining of order positions, 
wavelength calibration, were performed with the default settings of the UVES
pipeline. Then, before the order tracing and extraction of the spectra, the
cosmic rays in the low S/N science exposures were rejected by the
LACOSMIC (Laplacian Cosmic Ray Identification) routine from van Dokkum
(\cite{vando01}) which worked very well. This was necessary to avoid jumps
in the tracing as a result of cosmic rays close to the spectra.
For the further reduction, the inter-order background subtraction method
MEDIAN, the OPTIMAL extraction, flatfield correction in the extracted
pixel-order space, sky subtration, and merging of overlapping orders were
chosen. Also an error spectrum containing the standard deviation of the
reduced spectra was extracted for each object.

Since some spectra of the same object were observed at very different dates,
a correction for the heliocentric velocity was necessary before combination.
For UCD5, the heliocentric corrections differed by 35.08 km~s$^{-1}$ between
individual exposures. Since the instrumental spectral resolution is about
0.03$\AA$/pixel, a noticable shift in
wavelength (10\% $=$ 0.003$\AA$) already occurs for a velocity difference of
$\ge 0.15$ km~s$^{-1}$. The combination was done with the {\sc scombine} under
the {\sc noao.imred.echelle} package of {\sc IRAF}. Prior to averaging the
single spectra they were scaled by their mode, and pixel values deviating more
than 3 sigma from the average were rejected. In the same step the spectra were
rebinned to a resolution of 0.05 $\AA$/pixel.
The averaging and rebinning increased the S/N of a resolution element to
11-19 for the UCDs, depending on the object and the wavelength (see
Table~\ref{tab4}).

\begin{figure}
\psfig{figure=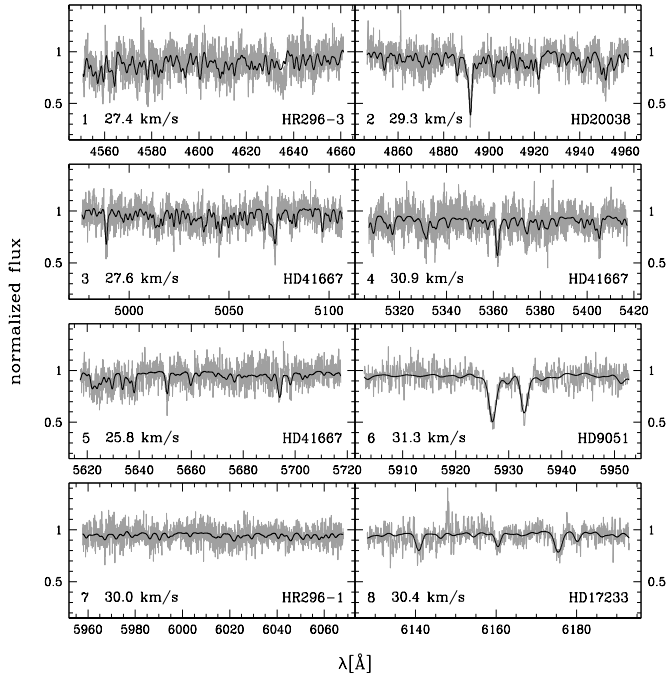,height=8.6cm,width=8.6cm
,bbllx=9mm,bblly=65mm,bburx=201mm,bbury=244mm}
\vspace{0.4cm}
\caption{\label{fig4} The direct fitting regions for the spectrum of UCD4
are shown (grey lines) together with the best fitting template (black lines).
The name of the template, the velocity dispersion of the gaussian kernel and
the number of the direct fitting region are indicated in each panel.
}
\end{figure}

\begin{figure}
\psfig{figure=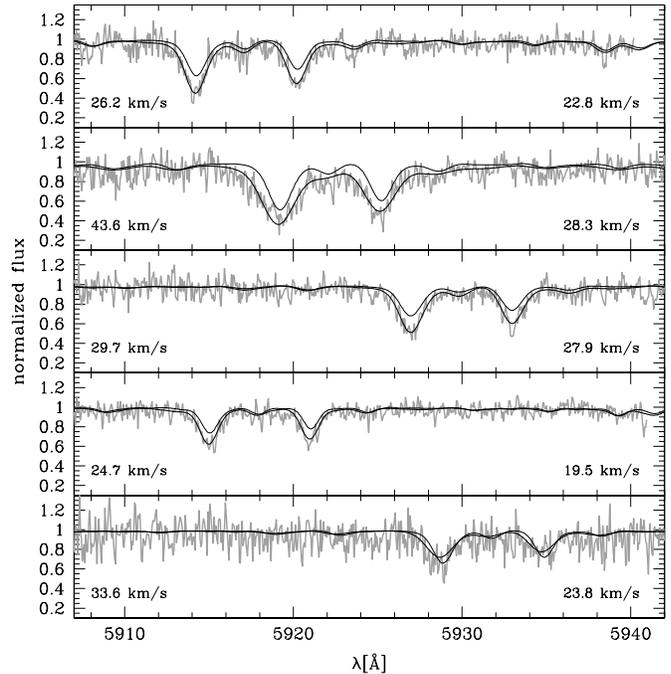,height=8.6cm,width=8.6cm
,bbllx=9mm,bblly=65mm,bburx=195mm,bbury=244mm}
\vspace{0.4cm}
\caption{\label{fig5} The Na doublet region for the five studied objects is
shown (UCD2, UCD3, UCD4, UCD5 and FCC303 from top to bottom).
The bold lines are averages of the three best-fitting templates for this region
(region 6, see Table~\ref{tab3}). The corresponding velocity dispersions are
given on the left. Thin lines represent the averages of the three best-fitting
templates in the regions 4, 5 and 7. Their velocity dispersions are given on
the right.
}
\end{figure}

\begin{table*}
\caption{\label{tab4} Results of velocity dispersion measurements in all 
direct fitting regions (same identification numbers as in Table~\ref{tab3}). 
On the left, the results from the blue part of the spectrum are shown,
and on the right the ones from the red part. The 
identification numbers of the three best fitting templates are given in 
brackets below each $\sigma_v$ value. The S/N are those for the combined
spectra at the given wavelength.}
\begin{tabular}{l@{\hspace{1mm}}|@{\hspace{1mm}}c@{\hspace{2mm}}c@{\hspace{2mm}}
c@{\hspace{2mm}}cc|@{\hspace{1mm}}c@{\hspace{2mm}}c@{\hspace{2mm}}
c@{\hspace{2mm}}c@{\hspace{2mm}}c@{\hspace{2mm}}cc}
\hline
 & & & & & & & & & & & \\[-3mm]
Name & S/N & \multicolumn{4}{c}{$\sigma_v$ [km s$^{-1}$]} & S/N &
 \multicolumn{6}{c}{$\sigma_v$ [km s$^{-1}$]} \\
 & & & & & & & & & & & & \\[-3mm]
 & 4900$\AA$ & 1 & 2 & 3 & 1+2+3 & 5900$\AA$ & 4 & 5 & 6 & 7 & 8 & 4+5+7 \\
\hline
 & & & & & & & & & & & & \\[-3mm]
UCD2 & 14.3 & 24.4$\pm$1.5 & 24.8$\pm$1.5 & 24.7$\pm$1.4 & 25.3$\pm$0.8 & 
 15.9 & 22.5$\pm$1.7 & 22.8$\pm$2.3 & 26.5$\pm$1.7 & 24.3$\pm$3.0 &
 19.4$\pm$2.0 & 24.4$\pm$1.2 \\
 & & (3,4,5) & (5,4,3) & (8,4,3) & (3,4,5) & & (6,5,4) & (3,5,8) & (8,2,5) & 
 (6,4,3) & (8,4,5) & (8,4,3) \\
 & & & & & & & & & & & & \\[-3mm]
UCD3 & 11.3 & 23.4$\pm$1.2 & 25.8$\pm$1.1 & 26.1$\pm$1.0 & 25.4$\pm$0.7 &
 13.0 & 29.2$\pm$1.4 & 29.6$\pm$2.0 & 44.3$\pm$1.6 & 26.8$\pm$3.0 & 
 32.3$\pm$2.6 & 30.0$\pm$1.1 \\
 & & (8,3,4) & (5,7,4) & (8,5,4) & (8,5,4) & & (7,5,3) & (5,8,3) & (2,8,5) &
 (6,3,4) & (8,5,4) & (7,3,4) \\
 & & & & & & & & & & & & \\[-3mm]
UCD4 & 14.9 & 27.6$\pm$1.9 & 30.1$\pm$1.7 & 27.5$\pm$1.6 & 29.3$\pm$1.0 &
 15.9 & 29.4$\pm$2.1 & 26.4$\pm$2.1 & 30.2$\pm$1.8 & 29.2$\pm$7.9 &  
 31.1$\pm$3.2 & 31.3$\pm$1.6 \\
 & & (8,4,3) & (4,5,3) & (4,5,2) & (3,4,5) & & (3,5,8) & (3,2,1) & (2,5,8) &
 (6,5) & (5,3,4) & (3,8,4) \\
 & & & & & & & & & & & & \\[-3mm]
UCD5 & 17.6 & 24.1$\pm$1.8 & 26.7$\pm$1.9 & 23.5$\pm$1.5 & 25.4$\pm$1.0 &
 19.1 & 20.1$\pm$1.7 & 19.0$\pm$1.8 & 23.8$\pm$1.9 & 20.3$\pm$2.7 &
 23.1$\pm$4.6 & 20.7$\pm$1.1 \\
 & & (3,4,5) & (3,4,2) & (3,4,2) & (3,4,5) & & (3,4,1) & (4,5,3) & (6,1,2) &
 (7,4,3) & (4,8,5) & (4,3,8) \\
 & & & & & & & & & & & & \\[-3mm]
FCC303 & 7.1 & 19.9$\pm$5.3 & 24.7$\pm$2.8 & 26.8$\pm$2.2 & 29.6$\pm$1.8 &
 9.3 & 24.4$\pm$2.6 & 23.3$\pm$2.7 & 34.0$\pm$4.0 & 6.5$\pm$3.9 &
 31.6$\pm$6.8 & 28.1$\pm$2.1 \\
 & & (3,2,1) & (4,2,5) & (2,4,5) & (3,4,2) & & (4,8,5) & (3,5,7) & (4,5,8) &
 (3,2,6) & (5,4,3) & (3,4,2) \\
\hline
\end{tabular}
\end{table*}

The standard star spectra have a high S/N ($\sim 50$-60) and were reduced with
the standard settings of the pipeline. They were rebinned to the same
resolution as the science spectra.

\section{Velocity dispersion measurements}

The kinematic analysis of the spectra was performed using van der Marel's
direct-fitting method (van der Marel \& Franx \cite{vdma93}). First, the
spectra were placed on a logarithmic wavelength scale and normalized. Then,
the standard star spectra were convolved with Gaussian velocity dispersion
profiles in the range 2 to 60 km~s$^{-1}$ and a step size of 1 km~s$^{-1}$.

For the comparison of the templates with the UCD spectra, wavelength
regions with prominent absorption features were selected. The restframe
wavelengths of the direct fitting regions and some prominent lines they cover
are summarized in Table~\ref{tab3} and are shown in Fig.~\ref{fig3}.
Additionally, direct fitting on a combination of three selected regions was
performed for the bluer part (regions 1, 2 and 3) as well as for the redder
part (regions, 4, 5 and 7) of the spectrum. All object spectra (UCDs and
FCC\,303) were fitted with all sets of smoothed template spectra (see
Table~\ref{tab2}). The best-fitting Gaussian velocity dispersion profile is
determined by $\chi^2$ minimization in pixel space. As an example, the best
fits to the spectrum of UCD4 for the eight direct fitting regions are shown in
Fig.~\ref{fig4}. In Table~\ref{tab4} the results for the three best-fitting
templates (average values) are given for all fitting regions. The
identification numbers of the three templates are given in brackets. In most
cases, the derived velocity dispersions of these templates are consistent
with each other within $\pm2$ km~s$^{-1}$. The template star HR296 was
observed in three different nights. The velocity dispersions as derived from
the three individual velocity dispersion fits with this star agree within
$\pm1$ km~s$^{-1}$ with each other in all cases. Thus no systematic effects
larger than $\pm2$ km~s$^{-1}$ either due to the use of different templates or
observations in different
nights are expected. The internal velocity dispersions of the stars are
expected to be below 3 km~s$^{-1}$, and thus do not affect the velocity
dispersion measurements. Note that one resolution element of 0.05$\AA$/pixel
corresponds to 2.5-3.5 km~s$^{-1}$.

The best-fitting templates are in most cases the ones with metallicities
between $-1$ and 0 dex. This should be expected since line index measurements
of the Fornax UCDs 2, 3 and 4 revealed iron abundances of [Fe/H]$= -0.90$,
$-$0.52 and $-$0.85 dex, respectively (Mieske et al. \cite{mies06}).

The scatter of velocity dispersion measurements between the different
direct fitting regions is very different for the bluer and redder part of the
spectrum. Whereas in the bluer part the scatter is of the order $\pm1.5$
km\,s$^{-1}$,
the regions 6 and 8 of the redder part, which include the Na doublet and a Ca
line, reveal large discrepancies towards high $\sigma_v$ values. This effect
is illustrated in Fig.~\ref{fig5} where the best fit to the Na doublet region
is compared to the expected fit in this region when taking the
average $\sigma_v$ of regions 4, 5 and 7. This discrepancy cannot
easily be explained. Probably abundance mismatches and/or stellar
population mixes play a role. The $\alpha$-abundances of UCDs in Virgo are
known to have super-solar values (Evstigneeva et al. \cite{evst06}). One might
also speculate whether the extra absorption might be caused by intra-cluster
material in the line of sight. Future studies shall clarify this issue.

Table~\ref{tab4} also shows that there seems to exist a slight systematic
effect between the velocity dispersion measurements in the bluer and redder
part of the spectrum (excluding regions 6 and 8) for some objects. UCD3 has
systematically higher $\sigma$ values in the red part as compared to the blue
part, whereas for UCD5 the contrary is the case. Also here, template
mismatches and contribution of stars from different evolutionary stages might
be the reason.

To define a final velocity dispersion for each object we consider
three cases of averaging the individual values: 1) simply the mean of all
fits; 2) the mean of all fits but with a 2-sigma clipping
of outliers; 3) mean excluding the Na and Ca region as well as the combined
regions. The results are listed in Table~\ref{tab5}. They are consistent
with each other within the errors. We adopt the average velocity dispersion
of case 2) as final value for further analyses.

\begin{table}
\caption{\label{tab5} Average values for radial velocity and velocity
dispersion measurements: (1) - mean of all fits; (2) - as (1), but with
2-sigma clipping; (3) - mean excluding Na and Ca region and combined region.
The values in bold face are those used for further analyses. The velocity
dispersion of UCD1 was derived by Drinkwater et al. (\cite{drin03a}) from Keck
spectroscopy.}
\begin{tabular}{lcccc}
\hline
 & & & & \\[-3mm]
Name & v$_{\rm helio}$ & $\sigma_v$ & {\bf $\sigma_v$} & $\sigma_v$ \\
 & [km s$^{-1}$] & [km s$^{-1}$] & [km s$^{-1}$] & [km s$^{-1}$] \\
 & & (1) & {\bf (2)} & (3) \\
\hline
 & & & & \\[-3mm]
UCD2 & 1205.7$\pm$0.8 & 24.4$\pm$1.6 & {\bf 24.7$\pm$1.0} & 24.1$\pm$1.0 \\
UCD3 & 1448.3$\pm$1.5 & 27.6$\pm$4.9 & {\bf 26.6$\pm$2.3} & 26.2$\pm$2.2 \\
UCD4 & 1839.8$\pm$3.8 & 29.3$\pm$1.7 & {\bf 29.2$\pm$1.5} & 28.4$\pm$2.0 \\
UCD5 & 1245.4$\pm$2.1 & 22.9$\pm$2.6 & {\bf 22.9$\pm$2.6} & 22.4$\pm$2.9 \\
FCC303 & 1928.6$\pm$3.1 & 26.1$\pm$5.6 & {\bf 27.1$\pm$3.1} & 23.1$\pm$6.0
\\[1mm]
UCD1 & ... & ... & 32.0$\pm$1.0 & ... \\[1mm]
\hline
\end{tabular}
\end{table}

For UCD1 no UVES spectroscopy is available. Yet its observed velocity
dispersion is known from Keck spectroscopy (Drinkwater et al. \cite{drin03a}).
It was derived from the CaT region using a cross-correlation method (Tonry
\& Davis \cite{tonr79}, as implemented in RVSAO/IRAF) and only one stellar
template (G6/G8IIw type). The determined value is given in Table~\ref{tab5}
and is used for the mass modelling (see next section). Note that in Drinkwater
et al. (\cite{drin03a}) also the velocity dispersions of UCD2 to UCD5 and
FCC303 were determined via the cross-correlation method. Since in this
preliminary study only the Na doublet region was used the obtained velocity
dispersions are 2-10 km\,s$^{-1}$ higher than the adopted values of the
present study (see discussion on the Na doublet above).

\section{Mass estimates and mass-to-light ratios}

The masses and mass-to-light ratios of the UCDs are important physical
parameters for the understanding of their origin. In particular, the
mass-to-light ratio ($M/L$) is an indicator for possibly existing dark matter
and/or violation of dynamical equilibrium.
If UCDs were the counterparts of globular clusters -- thus a single stellar
population without significant amounts of dark matter -- one would expect
$M/L$ values
as predicted by standard stellar population models (e.g. Bruzual \& Charlot
\cite{bruz03}, Maraston \cite{mara05}). If UCDs were of cosmological origin
-- thus formed in small, compact dark matter halos -- they might still be 
dominated by dark
matter and show a high $M/L$ value. Mass-to-light ratios that are larger than
expected from single stellar populations can, however, also be caused by
objects that are out of dynamical equilibrium, e.g. tidally disturbed stellar
systems (Fellhauer \& Kroupa \cite{fell06}) or by stellar populations that
formed with an unusual initial mass function.

In the following section we describe how the masses and mass-to-light ratios
were determined.

\subsection{Mass modelling}

The ingredients for the mass modelling of UCDs are their measured velocity
dispersions and their observed structural parameters. The light profiles
from Hubble Space Telescope imaging of several UCDs in Virgo and Fornax
are presented in Evstigneeva et al. (\cite{evst06}). It has been shown
that a simple King profile often is not the best choice to represent their
surface brightness profiles. Instead, Nuker laws or generalized King profiles
(with variable $\alpha$) fit the light profiles much better (see formulae in
Evstigneeva et al.).
Most of the mass estimators available in the literature, however, are based on
the assumption of a King model with exponent $\alpha =2$.

Not to be restricted only to King profiles a general approach has been 
developed. It calculates an objects' mass considering various representations 
of their light profiles. Thus, the following steps are performed:

\noindent
1. Fitting the observed luminosity profile by a given density law (Nuker, King, 
generalized King, Sersic or a two-component King+Sersic function).

\noindent
2. Deprojection of the fitted 2-dimensional surface density profile by means
of Abel's integral equation into a 3-dimensional density profile.

\noindent
3. Calculation of the cumulated mass function $M(<r)$ and the potential energy 
$\phi(r)$ from the 3-dimensional density profile. From these the energy 
distribution 
function $f(E)$ is then calculated with the help of eq.\ 4-140a from Binney \&
Tremaine (\cite{binn87}), assuming isotropic orbits for the stars.

\noindent
4. Creation of an $N$-body representation of the UCD using the de-projected 
density profile and the distribution function. For every model, 100.000 test 
particles were distributed. 

The modelling is based on the assumptions of spherical symmetry, an
underlying isotropic velocity distribution and no bias between mass and
light (i.e. mass segregation is neglected). 

Spherical symmetry is
justified since most UCDs are close to round with projected ellipticities
in the range $0.01<\epsilon<0.24$ (see Table 6, $\epsilon = 1-b/a$).
Their ellipticity distribution is consistant with that of Milky Way and 
NGC\,5128 globular clusters (Evstigneeva et al. \cite{evst06}). Concerning
the modelling, the potential is always more spherical than the surface 
density, so a small ellipticity does not have much influence on the 
measured velocity dispersion. 

The assumption of isotropy is justified by recent proper motion and radial 
velocity observations of $\omega$ Centauri, which show that the velocity 
dispersion of this cluster is close to isotropic inside two half-mass radii 
(van de Ven et al. \cite{vdve06}). Since $\omega$ Cen might be regarded as a
Local Group analogue for a UCD, the UCDs in our sample could have isotropic 
velocity dispersions as well. In addition, two-body relaxation will have 
isotropised the orbits of stars, at least in the inner parts of the studied 
UCDs. 

The assumption of mass 
following light is a critical issue. Whereas mass segregation of the stars 
can be neglected because of the long relaxation time of these massive objects,
an extended dark matter (DM) halo can not be excluded a priori. From the 
modelling point of view the addition of a DM halo is straightforward. Having, 
however, only one measured velocity dispersion (instead of a velocity 
dispersion profile), there is a large freedom in the scale length and total 
mass of the DM halo we model to match the observed velocity dispersion as 
well as the stellar density profile. In the cluster outskirts one could add 
nearly any amount of DM without changing the central velocity dispersion 
significantly. Not much can be learned from such an exercise. Therefore, we 
postpone the modelling of UCDs with extended dark matter halos to the future 
when spatially resolved velocity dispersion profiles become available.

\begin{figure}
\psfig{figure=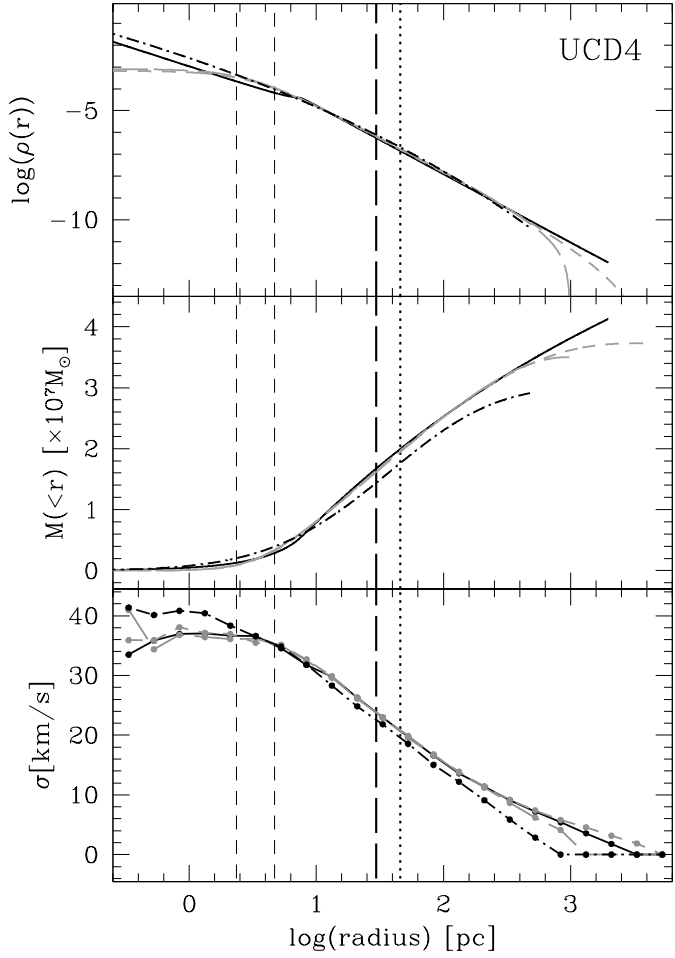,width=8.6cm
,bbllx=9mm,bblly=65mm,bburx=134mm,bbury=244mm}
\vspace{0.4cm}
\caption{\label{fig6} The model output of four different light profile
representations of UCD4 is shown; the line types are: black solid - Nuker law,
black dot-dashed - Sersic law, grey long dashed - King profile, grey short
dashed - generalized King profile.
The upper panel shows the 3-dimensional density distribution, the middle panel
the cumulative mass distribution, and the lower panel the projected velocity
dispersion profile of the $z$-component of the 3-dimensional velocity.
The vertical short dashed lines indicate the radii of 0.5 (left) and 1 pixel.
The dotted line is half the slit width, and the bold dashed line shows the
projected half-light radius of UCD4.
}
\end{figure}

\begin{figure}
\psfig{figure=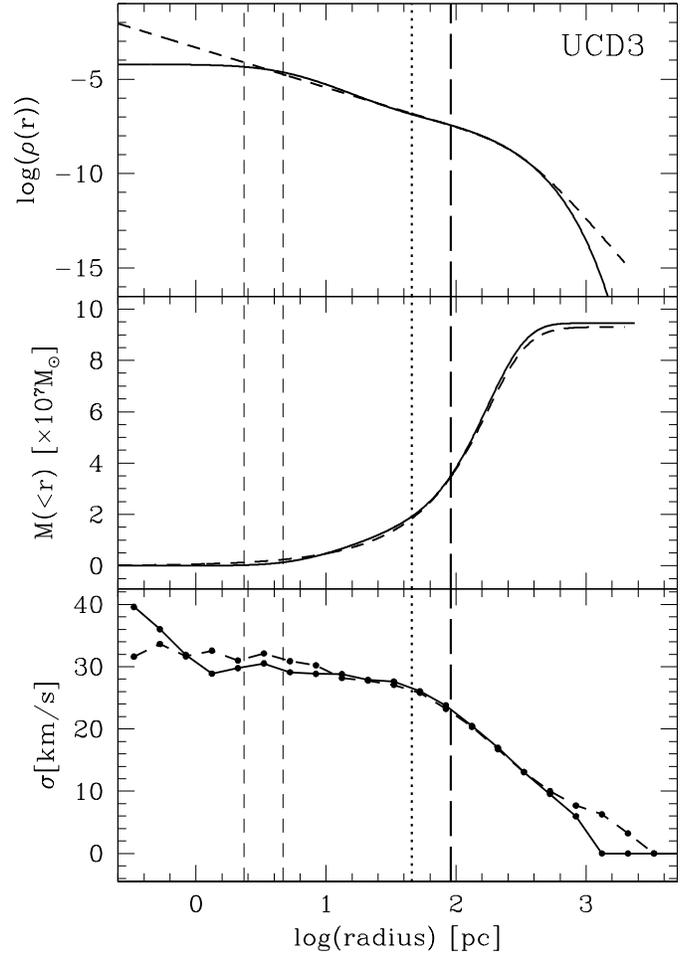,width=8.6cm
,bbllx=9mm,bblly=65mm,bburx=134mm,bbury=244mm}
\vspace{0.4cm}
\caption{\label{fig7} The model output of two different light profile
representations of UCD3 is shown; the line types are: solid - 2-component
King+Sersic profile, dashed - Nuker law.
The upper panel shows the 3-dimensional density distribution, the middle panel
the cumulative mass distribution, and the lower panel the projected velocity
dispersion profile of the $z$-component of the 3-dimensional velocity.
The vertical short dashed lines indicate the radii of 0.5 (left) and 1 pixel.
The dotted line is half the slit width, and the bold dashed line shows the
projected half-light radius of UCD3.
}
\end{figure}

Besides the projected profile parameters, the by us developed program expects 
as input the 
total mass of the stellar system and the number of test particles, and 
creates a list of $x$, $y$ and $z$ positions and $v_x$, $v_y$ and $v_z$ 
velocities for all particles that correspond to the specified structural 
parameters and the given mass. From this output file the central as well as 
global projected velocity dispersion can be calculated. The modelled projected
half-mass radius, $r_{h,p}$ is given as well for comparison with the observed 
value. In Fig.~\ref{fig6}, the 3-dimensional density distribution, the 
cumulative mass distribution and the projected velocity dispersion profile is 
shown in for four different model representations of UCD4 (see 
Table~\ref{tab6}). The same is shown for the Nuker and King+Sersic models
for UCD3 in Fig.~\ref{fig7}.

In case of the Nuker and Sersic functions, the models were truncated at large
radii to avoid the unphysical infinite extensions of UCD light profiles.
The truncation radius of the Nuker model was fixed at 2 kpc. Depending on the
steepness of the outer power law the total mass estimate can depend critically
on the exact value of the truncation radius. See, for example the rising
cumulative mass of UCD4 in Fig.~\ref{fig6} (middle panel).
The truncation radius of the Sersic
model was set to 20 times the effective radius, thus ranging between some
hundred parsecs and a few kiloparsecs for the UCDs in our sample. In some
cases, the adopted truncation radius might be too short leading to lower
values of the total mass as compared to other models (see Fig.~\ref{fig6},
middle panel).

\begin{figure}
\psfig{figure=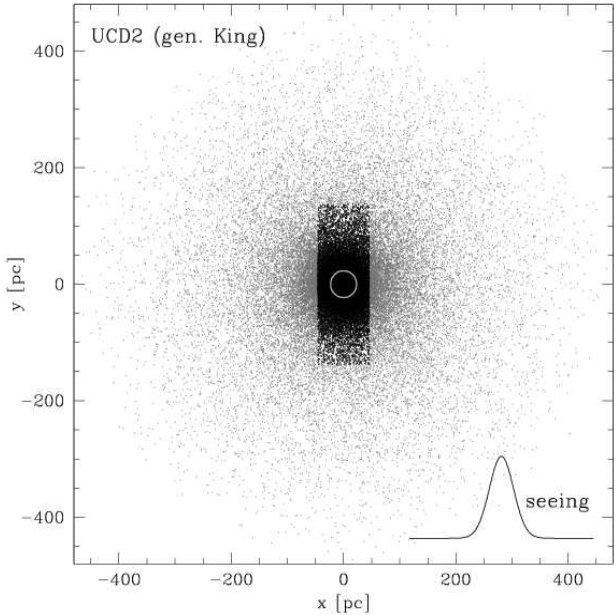,height=8.6cm,width=8.6cm
,bbllx=9mm,bblly=65mm,bburx=195mm,bbury=244mm}
\vspace{0.4cm}
\caption{\label{fig8} The $x$-$y$-plane of the modelled generalized King
representation of UCD2 is shown. Grey dots are all 100.000 test particles.
Black dots are those `stars' whose centres fall into the analysed slit area.
The grey circle indicates the projected half-light radius of UCD2.
The Gaussian in the lower right corner corresponds to the convolution kernel
used to estimate the fraction of light that falls into the slit
region for each `star'.
}
\end{figure}

The true tidal radii of the UCDs depend on their distances to the cluster 
center $R_G$ and the mass $m_c$ of the cluster (UCD). It can be 
estimated by the formula: $r_t = (Gm_c/2v_{\rm circ}^2)^{1/3} R_G^{2/3}$, 
where $v_{\rm circ}$ is the circular velocity of the cluster potential and
$G$ the gravitational constant. For the Fornax cluster, the circular velocity
is about 420 km\,s$^{-1}$ (e.g. Richtler et al. \cite{rich04}), and the 
projected distances of the UCDs are between 30 and 180 kpc.
The resulting lower limits (because of projected distances) 
of the tidal radii range between 1 and 2 kpc, agreeing with the 
adopted truncation radii.\\

After the creation of the UCD model the velocity dispersion as seen by an 
observer is calculated. To achieve this, the following steps are 
performed:

\noindent
1. All test particles are convolved with a Gaussian whose full-width at
half-maximum (FWHM) corresponds to the observed seeing.

\noindent
2. The fraction of the `light' (Gaussian) that falls into the
slit at the projected distance of the observed object (the size of the slit
in arcsec and the distance to the object in Mpc are input parameters) is
calculated.

\noindent
3. These fractions are used as weighting factors for the velocities. All
weighted velocities that fall into the slit region are then used to calculate
the `mimicked' observed velocity dispersion $\sigma_{\rm mod}$.

Of course, the total `true' mass of the modelled object, $M_{\rm true}$, that
corresponds to the observed velocity dispersion, $\sigma_{\rm obs}$ is not
known a priori. One has to start with a first guess of the total mass,
$M_{\rm guess}$, from which the `true' mass can be calculated as
$M_{\rm true} = M_{\rm guess}\cdot (\sigma_{\rm obs}/\sigma_{\rm mod})^2$.

\begin{figure}
\psfig{figure=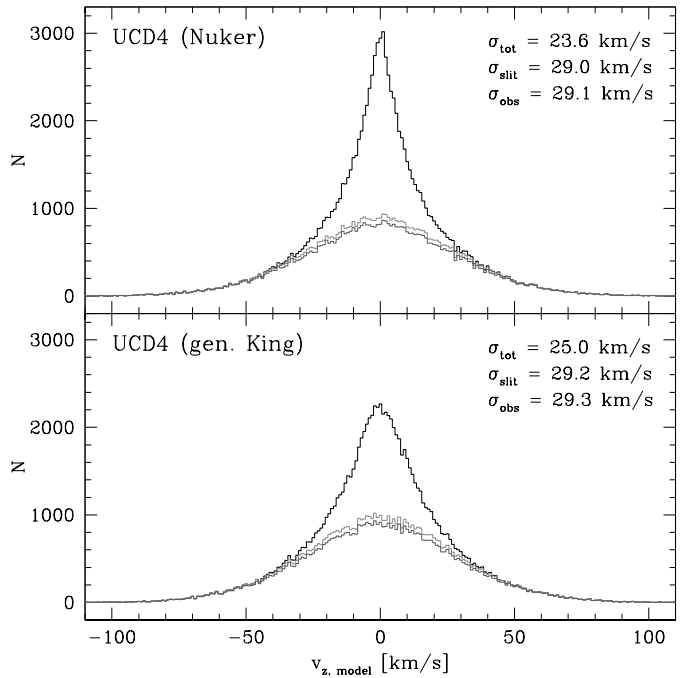,height=8.6cm,width=8.6cm
,bbllx=9mm,bblly=65mm,bburx=195mm,bbury=244mm}
\vspace{0.4cm}
\caption{\label{fig9} Velocity histograms of the $z$-component velocities
for the Nuker (top panel) and generalized King (lower panel) models of UCD4.
The bin width is 1 km\,s$^{-1}$. Black solid lines indicate the velocity
distributions of all test particles, the light grey lines those of the `stars'
with centres in the slit area, and the grey lines the luminosity weighted
number counts of those `stars' whose `light' falls into the slit area. The
corresponding velocity dispersions are given in the plots.
}
\end{figure}

As an example, in Fig.~\ref{fig8} the $x$-$y$-plane of the modelled
generalized King representation of UCD2 is shown. The black dots indicate
`stars' whose centres fall into the analysed slit area ($1\times3$ arcsec);
the Gaussian represents the point spread function profile of the `stars'
caused by the seeing conditions during the observations (see Table~\ref{tab1}).
In Fig.~\ref{fig9} the velocity histograms of all particles for two model
representations of UCD4 are shown. The particles in the histograms are the ones
in the slit area and the luminosity weighted number counts of those particles
whose `light' falls into the slit area.
One can see that the seeing hardly changes the velocity dispersion for
stars within the slit region. This is because most of the objects' light
is contained in the analysis area (e.g. the half-light radii of UCDs are
smaller than the slit width). Though, the total as well as central velocity
dispersion (see also Table~\ref{tab6}) can differ significantly from the
observed one due to the extended light profiles and central concentration
of the models, respectively.

\subsection{Dynamical masses}

The results of the modelled cluster masses and velocity dispersions for King,
generalized King, Sersic, Nuker and King+Sersic functions are presented in
Table~\ref{tab6}, together with the surface brightness profile model parameters
which were taken from Table 9 in Evstigneeva et al. (\cite{evst06}).

The errors of the modelled masses were estimated from the uncertainty in the
observed velocity dispersion, $\Delta\sigma$ and from typical uncertainties
of model parameters for surface brightness profiles (see Evstigneeva et al.
\cite{evst06}). On the one hand, maximum and minimum model masses were
estimated using the minimum and maximum observed velocity dispersion
$(\sigma_{\rm obs}+\Delta\sigma)$ and $(\sigma_{\rm obs}-\Delta\sigma)$,
respectively. The average of the
differences $(M_{\rm max}-M_{\rm true})$ and $(M_{\rm true}-M_{\rm min})$
defines the first mass error. On the other hand, models were created that
simulate $\sigma_{\rm obs}$ from the profile parameters plus/minus their
uncertainties. The maximum and minimum mass deviations from $M_{\rm true}$
define the second error. Both errors then were summed to derive the total
mass error.

\begin{table*}
\caption{\label{tab6} Mass modelling results for different light profile 
representations. The modelled projected half-light radius $r_{h,p}$, central 
velocity dispersion $\sigma_0$, global velocity dispersion $\sigma$, mass and 
mass-to-light ratio in the $V$-band are given.
The adopted surface brightness profile parameters are taken from Evstigneeva
et al. (\cite{evst06}). Radii ($R_b$, $R_e$, $R_c$ and $R_t$) are in pc; 
$\mu_{\rm 0,King}$ and $\mu_{e,\rm Sers.}$ are in mag\,arcsec$^{-2}$ in the
$V$ band.
In the last rows the results for the virial mass estimator are given. The
effective radius $R_{\rm eff}$ was taken from Table 10 of Evstigneeva et al. 
(\cite{evst06}), the global velocity dispersion from the generalized King 
models and King + Sersic models in two-component fits case. Also the
ellipticies ($\epsilon = 1-b/a$) from Evstigneeva et al. are shown for
reference. The values in bold face were adopted for further analyses.}
\begin{tabular}{p{1.7cm}ccccccccccc}
\hline
 & & & & & & & & & & & \\[-3mm]
Model & & & & & & & $r_{h,p}$ & $\sigma_0$ & $\sigma$ & $M_{\rm mod}$ & 
 $M_{\rm mod}/L_V$ \\
 & & & & & & & [pc] & [km~s$^{-1}$] & [km~s$^{-1}$] & $\times10^7M_\odot$ & \\
\hline
 & & & & & & & & & & & \\[-3mm]
{\bf Nuker:} & $R_b$ & $\alpha$ & $\beta$ & $\gamma$ & & & & & & & \\
\hline
 & & & & & & & & & & & \\[-3mm]
UCD1  & 36.7 & 0.34 & 3.03 & 0.91 & ... & ... & $39.1\pm3.0$ & $51.4\pm3.0$ & 
 $25.2\pm2.0$ & $2.59\pm0.32$ & $4.03\pm0.53$ \\
UCD2  &  4.3 & 0.73 & 2.38 & 0.58 & ... & ... & $39.4\pm3.0$ & $33.6\pm1.0$ & 
 $20.4\pm2.0$ & $2.39\pm0.38$ & $3.45\pm0.59$ \\
UCD3  & 318.7 & 1.90 & 7.04 & 1.10 & ... & ... & $89.2\pm8.0$ & $33.0\pm1.8$ & 
 $22.7\pm3.3$ & $9.06\pm2.25$ & $3.95\pm1.00$ \\
UCD4  &  7.3 & 20.00 & 2.13 & 0.88 & ... & ... & $50.4\pm4.0$ & $36.4\pm1.2$ &
 $23.6\pm3.6$ & $4.13\pm1.02$ & $5.06\pm1.31$ \\
\hline
 & & & & & & & & & & & \\[-3mm]
{\bf Sersic:} & $R_e$ & $n$ & & & & & & & & & \\
\hline
 & & & & & & & & & & & \\[-3mm]
UCD1  & 36.9 & 9.9 & ... & ... & ... & ... & $33.1\pm3.5$ & $47.3\pm4.5$ & 
 $26.1\pm3.0$ & $2.59\pm0.48$ & $4.03\pm0.77$ \\
UCD2  & 26.6 & 6.8 & ... & ... & ... & ... & $25.7\pm2.5$ & $35.5\pm0.8$ & 
 $21.8\pm2.5$ & $1.93\pm0.39$ & $2.79\pm0.58$ \\
UCD4  & 24.1 & 5.5 & ... & ... & ... & ... & $23.8\pm2.5$ & $40.0\pm1.1$ & 
 $26.3\pm4.8$ & $2.91\pm0.96$ & $3.56\pm1.22$ \\[1mm]
\hline
 & & & & & & & & & & & \\[-3mm]
{\bf King, $\alpha=2$:} & $R_c$ & $R_t$ & $c$ & & & & & & & & \\
\hline
 & & & & & & & & & & & \\[-3mm]
UCD1  & 1.8 & 5761.2 & 3.51 & ... & ... & ... & $49.2\pm3.5$ & $40.7\pm1.0$ &
 $24.3\pm4.0$ & $4.17\pm1.04$ & $6.49\pm1.61$ \\
UCD2  & 2.3 & 1457.5 & 2.80 & ... & ... & ... & $28.3\pm1.5$ & $31.5\pm0.8$ &
 $21.2\pm3.0$ & $2.40\pm0.58$ & $3.47\pm0.86$ \\
UCD4  & 2.8 & 987.6 & 2.55 & ... & ... & ... & $26.1\pm1.3$ & $36.4\pm1.3$ &
 $25.6\pm5.1$ & $3.50\pm1.22$ & $4.29\pm1.55$ \\[1mm]
\hline
 & & & & & & & & & & & \\[-3mm]
{\bf King, var. $\alpha$:} & $R_c$ & $R_t$ & $c$ & $\alpha$ & & & & & & & \\
\hline
 & & & & & & & & & & & \\[-3mm]
UCD1  & 1.8 & 378.0 & 2.32 & 0.74 & ... & ... & $22.4\pm1.0$ & 
 {\bf 41.3$\pm$1.0} & {\bf 27.1$\pm$1.8} & {\bf 3.21$\pm$0.36} & 
 {\bf 4.99$\pm$0.60} \\
UCD2  & 2.2 & 487.1 & 2.35 & 1.23 & ... & ... & $23.2\pm1.0$ & 
 {\bf 31.3$\pm$0.6} & {\bf 21.6$\pm$1.8} & {\bf 2.18$\pm$0.31} & 
 {\bf 3.15$\pm$0.49} \\
UCD4  & 3.0 & 4501.2 & 3.18 & 3.32 & ... & ... & $29.6\pm2.0$ & 
 {\bf 37.3$\pm$0.6} & {\bf 25.0$\pm$3.4} & {\bf 3.73$\pm$0.86} & 
 {\bf 4.57$\pm$1.11} \\[1mm]
\hline
 & & & & & & & & & & & \\[-3mm]
{\bf King+Sersic:} & $R_c$ & $R_t$ & $\mu_{\rm 0,King}$ & $R_e$ & $n$ & 
$\mu_{e,\rm Sers.}$ & & & & & \\
\hline
 & & & & & & & & & & & \\[-3mm]
UCD3  & 3.6 & 119.2 & 15.74 & 106.6 & 1.34 & 21.13 & $89.9\pm6.0$ & 
 {\bf 29.3$\pm$1.2} & {\bf 22.8$\pm$3.1} & {\bf 9.45$\pm$2.20} & 
 {\bf 4.13$\pm$0.98} \\
UCD5  & 4.0 & 30.7 & 15.13 & 134.5 & 6.9 & 24.04 & $30.0\pm2.5$ & 
 {\bf 28.7$\pm$0.8} & {\bf 18.7$\pm$3.2} & {\bf 1.80$\pm$0.50} & 
 {\bf 3.37$\pm$0.85} \\
FCC303 & 1.2 & 2403.9 & 13.32 & 696.3 & 0.6 & 22.81 & $668.3\pm12.$ & 
 {\bf 37.8$\pm$3.5} & {\bf 16.5$\pm$3.9} & {\bf 33.15$\pm$9.53} & 
 {\bf 2.47$\pm$0.72} \\[1mm]
\hline
 & & & & & & & & & & & \\[-3mm]
 \multicolumn{3}{l}{\bf Virial mass estimator:} & & & & $\epsilon$ & & 
 $R_{\rm eff}$ & $\sigma$ & $M_{\rm vir}$ & $M_{\rm vir}/L_V$ \\
\hline
 & & & & & & & & & & & \\[-3mm]
UCD1   & ... & ... & ... & ... & ... & 0.18 & ... & 22.4  & $27.1\pm1.8$ &
 $3.73\pm0.52$ & $5.70\pm0.84$ \\
UCD2   & ... & ... & ... & ... & ... & 0.01 & ... & 23.1  & $21.6\pm1.8$ &
 $2.44\pm0.42$ & $3.47\pm0.62$ \\
UCD3   & ... & ... & ... & ... & ... & 0.03 & ... & 92.2  & $22.8\pm3.1$ &
 $10.87\pm3.04$ & $4.66\pm1.32$ \\
UCD4   & ... & ... & ... & ... & ... & 0.05 & ... & 29.5  & $25.0\pm3.4$ &
 $4.18\pm1.17$ & $5.03\pm1.43$ \\
UCD5   & ... & ... & ... & ... & ... & 0.24 & ... & 31.2  & $18.7\pm3.2$ &
 $2.47\pm0.87$ & $4.54\pm1.61$ \\
FCC303 & ... & ... & ... & ... & ... & 0.03 & ... & 660.0 & $16.5\pm3.9$ &
 $40.74\pm19.3$ & $2.98\pm1.42$ \\
\hline
\end{tabular}
\end{table*}

The mass-to-light ratio errors were propagated from the mass errors
and the uncertainty in the luminosity (assumed to be 0.05 mag in the absolute
magnitude). The luminosities were derived from the apparent, extinction
corrected $V$ magnitude as
given in Table~\ref{tab1}, a distance modulus to Fornax of $(m-M)_V=31.39$ mag
(Freedman et al. \cite{freed01}) and a solar absolute $V$ magnitude of
$M_{V,\odot} = 4.85$ mag.

The error of the central velocity dispersion was estimated from the
observational error plus the scatter of modelled velocity dispersions in
annuli of 0.5 parsec within the central 5 pc for each object. The error of
the global velocity dispersion is the sum of the observational error and
the uncertainties as propagated from the mass error of the profile
parameters.

The masses and $M/L$ values of the different models in general agree with
each other within the errors. In Fig.~\ref{fig10}, the luminosity versus mass
plane is shown for all model results and the virial mass estimate (see below).
The mass-to-light ratios of most UCDs scatter around $M/L_V=4$. 
On average, the masses of the Sersic profile
are slightly lower than those derived from the other profiles, mainly caused
by the adopted truncation radius (see discussion above). The masses of the
Nuker profile are sensible to the slope of the outer power law and the
adopted truncation radius. The shallow outer slope of UCD4 causes a high mass
estimate, whereas the rather steep outer slope of UCD1 results in a low mass
estimate as compared to the other models.
Besides the uncertainties in the truncation radius, Nuker and Sersic mass 
estimates suffer from the uncertainties in the central parts of the 
profile. The cuspy nature of the density profile depends on the light 
distribution within the central 2-4 pixels, an area for which it is difficult 
to make a model
independent of the assumed fitting function (see Fig.~3 in Evstigneeva et al.
\cite{evst06}). Most sensitively, the central velocity dispersion $\sigma_0$
depends on the model shape within the innermost pixels. The derived $\sigma_0$
values for Nuker or Sersic models are 3-10 km\,s$^{-1}$ higher than those
for the King models. This has to be taken into account when using central
velocity dispersions in fundamental plane plots.

\begin{figure}
\psfig{figure=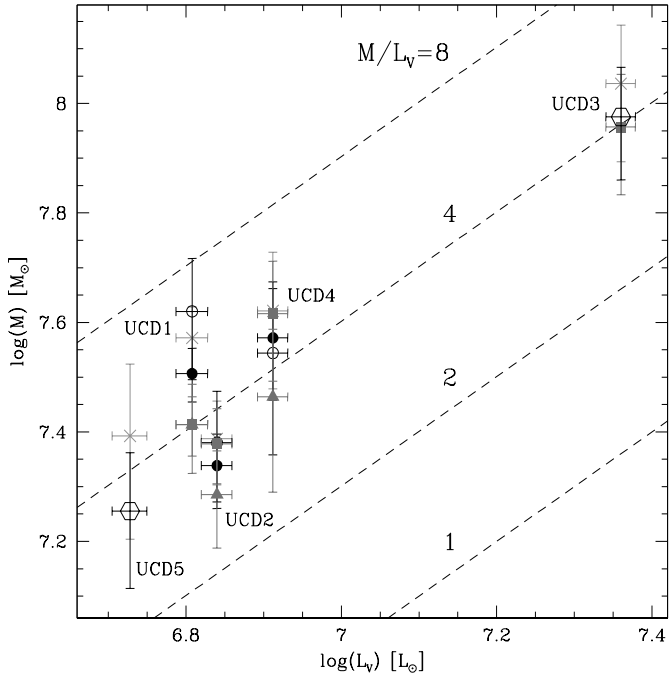,height=8.6cm,width=8.6cm
,bbllx=9mm,bblly=65mm,bburx=195mm,bbury=244mm}
\vspace{0.4cm}
\caption{\label{fig10} Luminosity vs. mass plot
for five Fornax UCDs (the data for UCD1 were taken from Drinkwater et al.
(\cite{drin03a}). The symbols indicate the following models: grey open circles
- King, filled circles - generalized King, grey triangles - Sersic; grey
squares - Nuker, open hexagons - King+Sersic, and crosses - virial mass
estimator. Lines of constant mass-to-light ratio are shown as indicated.
}
\end{figure}

The most stable results for one-component profile fits are those of the
generalized King model because they do not suffer from the uncertain profile
shape in the inner pixels (like Nuker and Sersic) and fit the outer regions
much better than the `normal' King models. Concerning two-component profiles, 
King core+Sersic halo models are the most reliable. Therefore, the fit values
of these models were adopted for further analyses (see numbers in bold face 
in Table~\ref{tab6}).

The results of our dynamical modelling were also compared with the virial
mass estimator by Spitzer (\cite{spitz87}), $M_{\rm vir} \approx 9.75 (R_{\rm
eff} \sigma^2)/G$, where $\sigma$ is the global projected velocity dispersion
and $R_{\rm eff}$ the half-light radius. This method also assumes a constant 
$M/L$ ratio as function of radius and an isotropic velocity distribution. The 
results are shown in Table~\ref{tab6}. The errors in the virial mass estimates
were propagated from the errors in $R_{\rm eff}$ and $\sigma$.
The virial and modelled masses agree within the errors.

\subsection{Comparison with masses from stellar population models}

In turn the masses and mass-to-light ratios as derived from dynamical
modelling are compared with the $M/L$ values expected from stellar 
population models in order to see whether UCDs can be explained by pure 
stellar systems or whether they would need dark matter within the central
1-3 half-mass radii to account for missing mass.

\begin{figure}
\psfig{figure=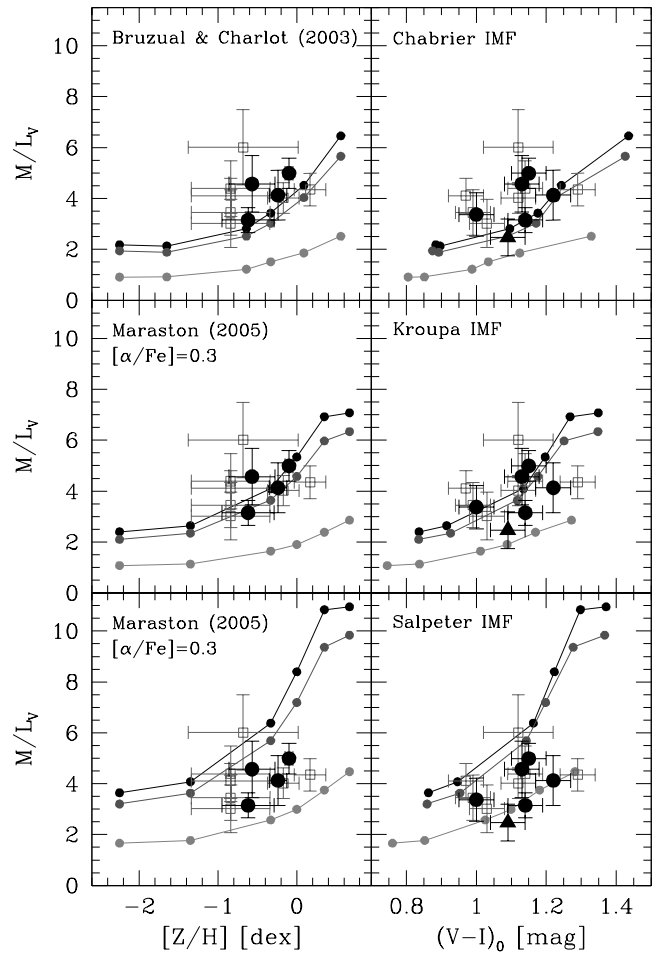,width=8.6cm
,bbllx=9mm,bblly=65mm,bburx=134mm,bbury=244mm}
\vspace{0.4cm}
\caption{\label{fig11} The dynamical $M/L_V$ ratios of Fornax UCDs (filled 
circles) and the nucleus of FCC303 (filled triangle) are compared to expected 
$M/L_V$ values from stellar population models. Also shown are data for Virgo
UCDs (open squares from Evstigneeva et al. (\cite{evst06}). In the left panels
$M/L_V$ is plotted versus
metallicity [Z/H], in the right panels versus the colour $(V-I)_0$. The
references to the models and the adopted IMF are indicated in the panels.
The three lines in each plot represent ages of 15, 13 and 5 Gyr from top to
botton, respectively. For the Fornax UCDs an $\alpha$ abundance of 
[$\alpha$/Fe]$=0.3$ was adopted to derive their [Z/H] values. Note, however,
that their $\alpha$ abundance might be different (solar) if they possess a
significant contribution of intermediate age stellar populations.
}
\end{figure}

There are stellar population models available from various groups. Here we
present the comparison with Bruzual \& Charlot (\cite{bruz03}) and Maraston 
(\cite{mara05}) models. Mass-to-light ratios are given for single stellar 
populations (SSPs) of various ages and metallicities. The metallicity in the 
Maraston models is a total metallicity [Z/H] including possible [$\alpha$/Fe]
overabundances. [Z/H] and the iron abundance [Fe/H] are related as following: 
[Fe/H] $=$ [Z/H] $- 0.94\cdot$[$\alpha$/Fe] (Thomas et al. \cite{thoma03}). 
[Z/H] is in agreement with the Zinn \& West (\cite{zinn84}) scale.

In Fig.~\ref{fig11}, the dynamical masses of Fornax UCDs and the nucleus of
FCC\,303 are compared
to stellar population models in the [Z/H] versus $M/L_V$ and $(V-I)$ versus 
$M/L_V$ plane. For comparison we also include data for seven of the brightest
Virgo UCDs (Evstigneeva
et al. \cite{evst06}). Three sets of models with different assumption for the 
initial mass function (IMF) are presented. The Bruzual \& Charlot models with a
Chabrier IMF, and the Maraston models with a Kroupa as well as a Salpeter IMF.
These IMFs were chosen to illustrate the wide range of predictions from various
SSP models. The colours of the Virgo UCDs were taken from Evstigneeva et al. 
(\cite{evst06}). The ages, metallicities and $\alpha$ abundances were 
estimated from diagnostic plots of line indices in Evstigneeva et al. (Figs. 
11 \& 12). Most Virgo UCDs are old (8 to 15 Gyr), and most
of them show super-solar [$\alpha$/Fe] values around 0.3 dex. 
The iron abundances and colours of the Fornax UCDs were taken from Mieske et 
al. (\cite{mies06}). No age and $\alpha$-abundance estimates are published
for them. The relatively high Balmer line H$\beta$ indices of
UCD3 and UCD4 (see Table 1 in Mieske et al. \cite{mies06}) might point to a 
contribution of young to intermediate age stellar populations within these 
UCDs. Note that this statement has to be taken with caution since no proper
calibration of the line indices was performed in Mieske at al.'s paper.

Fig.~\ref{fig11} shows that the dynamical masses of all UCDs and the nucleus
can be reconciled 
with stellar population models if taking the `right' set of models. 
The Bruzual \& Charlot models with a Chabrier IMF tend to predict $M/L_V$ 
values that are lower than the measured ones. Taking these models at face value
one might conclude that most UCDs need some amount of dark matter to explain
their higher $M/L_V$ values.
When taking Maraston models with a Salpeter IMF, however, even the extreme case
of Strom417 (a low mass UCD or massive GC) with a dynamical $M/L_V$ value 
around 6 (Evstigneeva et al. 
\cite{evst06}, Ha\c{s}egan et al. \cite{hase05}) can be explained by an old 
stellar population. If these models are correct for all UCDs,
many of them are best fitted by intermediate age (5-8 Gyr) stellar populations.
In the Fornax cluster, UCD2, UCD3 and the nucleus of FCC303 would be candidates
for young ages, whereas UCD1, UCD4 and UCD5 still are compatible with old 
stellar populations. In the Virgo cluster, VUCD3, VUCD5 and VUCD6 are possibly
younger than the other UCDs. This, however, contradicts the findings by
Evstigneeva et al. (\cite{evst06}) who derived old ages for basically all UCDs
in Virgo.

It is clear from the discussion above that to decide whether dynamical masses
of UCDs are really consistent with a pure stellar population is not
straightforward.
On the one hand, more accurate ages, metallicities and 
$\alpha$-abundances are needed to better constrain the right choice of the 
stellar population models (i.e. to constrain the IMF). On the other
hand, there still is the freedom of adding an extended dark matter halo when
modelling the UCDs for deriving their masses. In this case, also Bruzual \& 
Charlot models with a Chabrier IMF could be the `right' choice.

The most important result that comes out from our comparison between
dynamical and stellar population model masses is that all UCDs and the nucleus
of FCC\,303 can, in principle, be explained by a pure stellar 
population. Dark matter is not mandatory
for any of the objects within 1-3 half-mass radii. As mentioned before,
it can not be
ruled out, however, that UCDs are dominated by dark matter at large radii
where high signal-to-noise spectra hardly can be obtained due to the very
low surface brightness.

\section{Summary and Conclusions}

In this paper we have presented the spectroscopic analysis of four 
ultra-compact dwarf galaxies (UCDs) and one nucleus of a dwarf elliptical.
These UCDs are among the five brightest in the central region of the Fornax
cluster (Hilker et al. \cite{hilk99c}, Drinkwater et al. \cite{drin03a}).
The analysis is based on high resolution spectra obtained with the instrument 
UVES mounted in the VLT at El Paranal, Chile. Velocity dispersions were derived
from selected wavelength regions using the direct-fitting method by van der
Marel (van der Marel \& Franx \cite{vdma93}). 
To derive the masses of the UCDs a new modelling program has been
developed that allows to choose between different representations of the
surface brightness profile (i.e. Nuker, Sersic or King laws) and corrects
the observed velocity dispersions for observational parameters (i.e. seeing,
slit size). The light profile parameters of the UCDs and the nucleus of 
FCC\,303 were obtained from Hubble
Space Telescope imaging (see the accompanying paper by Evstigneeva et al. 
\cite{evst06}). The derived dynamical masses then have been compared to those
expected from stellar population models.
The main results of our analysis are as follows:\\

(i) The observed velocity dispersions of the Fornax UCDs range between 22 and
30 km\,s$^{-1}$,
slightly smaller than those of Virgo UCDs (Evstigneeva et al. \cite{evst06}) 
and DGTOs (Ha\c{s}egan et al. \cite{hase05}). The observed velocity dispersions
of the Na doublet regions are systematically higher than those of the other 
regions, probably reflecting the contribution of stars from different
evolutionary stages.

(ii) The generalized King models for the one-component fits and King+Sersic 
models for the two-component fits were adopted to determine the masses of the 
UCDs,
which range between 1.8 and $9.5\times10^7M_{\odot}$. The corresponding central
and global projected velocity dispersions are in the range 29 to 41 and 19 to
27 km\,s$^{-1}$, respectively. The masses as derived from Nuker and Sersic 
models depend on the adopted truncation radius and the slope of the outer power
law (Nuker). Truncation radii that are too small lead to underestimated masses 
(e.g. Sersic model of UCD4) and flat outer power laws to a steeply increasing
cumulative mass profile 
(e.g. Nuker model of UCD4). The central velocity dispersion mainly depends on 
the shape of the light profile within the innermost parsecs. If the cuspy 
behaviour of some UCDs is true, the central velocity dispersion would be up to
10 km\,s$^{-1}$ higher than derived from generalized King models.

(iii) The mass-to-light ratios $M/L_V$ of the Fornax UCDs range between 3 and 
5. These values are compatible with predictions from stellar population models,
especially those from Maraston (\cite{mara05}) assuming a Kroupa or Salpeter
IMF. A Salpeter IMF, however, would imply that about half of the Fornax UCDs 
would be dominated by intermediate age populations of 4 to 8 Gyr. The high
Balmer H$\beta$ indices of three UCDs in Fornax is consistent with this 
interpretation (Mieske et al. \cite{mies06}). Also the $M/L_V$ values of all 
Virgo UCDs can be explained by stellar population models (see Evstigneeva et 
al. \cite{evst06}). No dark matter component is needed for UCDs whithin
1-3 half-mass radii. One can, however, not rule out that UCDs might be 
dominated by dark matter in the outer parts.

(iv) The mass-to-light ratio $M/L_V$ of the nucleus of FCC\,303 is 2.5, 
entirely compatible with a pure stellar population and similar to those of
other dwarf elliptical nuclei (Geha et al. \cite{geha02}).\\ 

We conclude that ultra-compact dwarf galaxies most probably are the result of 
cluster formation processes (i.e. like large globular clusters, assembled star
cluster complexes, nuclear star clusters) rather than being genuine 
cosmological sub-structures themselves (i.e. compact galaxies formed in small,
compact dark matter halos). 

Massive star clusters (and
star cluster complexes) nowadays form in interacting galaxies like the 
Antennae (e.g. Whitmore et al. \cite{whitm99}). It has been shown that the
most massive and compact young massive clusters (YMCs) and star cluster 
complexes can survive their `infant mortality' (e.g. Fellhauer \& Kroupa
\cite{fell02a}, Bastian et al. \cite{bast06a}).
The most massive surviving YMCs have masses comparable to those of the
UCDs presented in this paper (NGC7252:W3, Maraston et al. \cite{mara04};
NGC7252:W30, NGC1316:G114, Bastian et al. \cite{bast06b}).
It is reasonable to assume that similar massive clusters have formed in
merger events of the early epochs of galaxy cluster formation. Some of them 
might have survived the disruption and merging of galaxies during the galaxy 
cluster formation process as free-floating `UCDs'.

But also the threshing of nucleated dwarf galaxies can not be ruled out 
definitely as formation process of UCDs. Nuclear star clusters as massive
as UCDs were able to form either by the merging of globular clusters through
tidal friction in early-type dwarf galaxies (e.g. Oh \& Lin \cite{oh00}, Lotz 
et al. \cite{lotz01}) or through successive central starbursts in late-type 
bulge-less spirals. During the dynamical evolution of galaxy clusters many of
these low mass nucleated galaxies may be disrupted (e.g. Moore et al. 
\cite{moor99a}). Their debris then is distributed in the intra-cluster 
medium, building up the extended cD halos of central galaxies (e.g. Hilker et 
al. \cite{hilk99b}). Their globular clusters and nuclei are expected to 
survive as intra-cluster stellar populations, among them the present-day 
dark matter-free `UCDs'.

In order to differentiate between the formation of UCDs via super-star cluster
formation in starbursts and the formation via tidal disruption of nucleated
(dwarf) galaxies detailed analysis of the chemical composition and age 
structure of the stellar population in UCDs might help. A super-solar 
[$\alpha$/Fe] abundance (as suggested for most of the Virgo UCDs, Evstigneeva
et al. \cite{evst06}) would point to a rapid, short-timescale formation 
process which is typical for old globular clusters. A galaxy nucleus would 
only show a super-solar [$\alpha$/Fe] value if its star formation was 
truncated soon after its initial burst or if it would be composed of merged
GCs which themselves have super-solar $\alpha$ abundances.
However, present-day nuclei of dwarf ellipticals seemed to have
formed in a rather continuous way since their $\alpha$-abundances are enriched
to solar values (Geha et al. \cite{geha03}). It has to be seen whether the 
$\alpha$-abundances of the Fornax UCDs are the same as for the Virgo UCDs.
Intermediate age populations in UCDs would indicate that some Gyr ago gas
still was present in UCDs, constraining the threshing timescale and/or the 
time of the last major star formation event.
Note that there are some hints that Fornax UCDs might be different to Virgo 
UCDs (e.g. Mieske et al. \cite{mies06}). Maybe also the environment plays a 
role in their formation processes. 

Independent of which scenario is right it is interesting to note that star
clusters more massive than about $5\times10^6M_{\odot}$ seem to behave 
different than less massive `ordinary' GCs. Not only the scaling and 
fundamental plane relations of massive clusters start to deviate from those of
normal GCs (see results and discussions in Kissler-Patig et al. \cite{kiss06} 
and Evstigneeva et al. \cite{evst06}), but also their stellar populations are
getting more complex. Massive clusters do not possess a single stellar 
population, but are composed of sub-populations if different metallicities
and/or ages. The best example is the most massive globular cluster 
in our Milky Way, $\omega$ Centauri, whose stellar populations revealed a
broad metallicity distribution as well as an extended star formation history
(e.g. Hilker \& Richtler \cite{hilk00b}, Hilker et al. \cite{hilk04a}). 
Also G1, one of the most massive GCs of M31 is suspected to contain multiple
stellar populations (e.g. Meylan et al. \cite{meyl01}).
It would be interesting to know whether this is true for all massive
clusters, including UCDs.

Although our study has shown that UCDs most probably can be explained by pure
stellar populations, more constraints from observations are needed in order 
to get further insights into their nature. High signal-to-noise spectra for 
the Fornax UCDs would be useful to derive more accurate chemical abundances 
and age estimates from appropriate line indices. This should clarify whether 
they are enhanced in $\alpha$ elements and whether they posses intermediate
age populations. 
Also high resolution spectroscopy (to obtain internal velocity dispersions)
of more UCDs and galactic nuclei is needed to fill the parameter space
(fundamental plane) in the interface between globular clusters and dwarf
galaxies. The most extended UCDs might be targets for spatially resolved high 
resolution spectroscopy. This would show whether our modelled velocity
dispersion profiles are real or whether one would need
more sophisticated modelling to comply with possible orbital anisotropies 
and/or violation of the virial equilibrium through tidal effects. 
Spatially resolved velocity dispersion profiles also will allow to constrain
the shape and mass of extended dark matter halos around UCDs if they existed.

\acknowledgements
LI acknowledges support from the Fondap ``Centre of Astrophysics''.
This work has been supported by the United States National Science Foundation
grant No.~0407445 and by NASA through grants GO-8685 and GO-10137 from the
Space Telescope Science Institute, which is operated by AURA, Inc., under NASA
contract NAS5-26555. Part of the work reported here was done at the Institute
of Geophysics and Planetary Physics, under the auspices of the U.S. Department
of Energy by Lawrence Livermore National Laboratory under contract 
No.~W-7405-Eng-48.

\end{document}